\documentstyle[preprint,aps,eqsecnum,epsf,floats]{revtex}
\begin{document}
\tighten

\def\bfl{{\bbox \ell}}
\def\bull{\vrule height .9ex width .8ex depth -.1ex}
\def\MeV{{\rm MeV}}
\def\GeV{{\rm GeV}}
\def\Tr{{\rm Tr\,}}
\def\nrcpt{NR\raise.4ex\hbox{$\chi$}PT\ }
\def\ket#1{\vert#1\rangle}
\def\bra#1{\langle#1\vert}
\def\ltap{\ \raise.3ex\hbox{$<$\kern-.75em\lower1ex\hbox{$\sim$}}\ }
\def\gtap{\ \raise.3ex\hbox{$>$\kern-.75em\lower1ex\hbox{$\sim$}}\ }
\def\abs#1{\left| #1\right|}
\def\CA{{\cal A}}
\def\CC{{\cal C}}
\def\CD{{\cal D}}
\def\CE{{\cal E}}
\def\CL{{\cal L}}
\def\CO{{\cal O}}
\def\CZ{{\cal Z}}
\def\bvert{\Bigl\vert\Bigr.}
\def\pds{{\it PDS}\ }
\def\ms{MS}
\def\ddq{{{\rm d}^dq \over (2\pi)^d}\,}
\def\ddqm{{{\rm d}^{d-1}{\bf q} \over (2\pi)^{d-1}}\,}
\def\bfq{{\bf q}}
\def\bfk{{\bf k}}
\def\bfp{{\bf p}}
\def\bfpp{{\bf p '}}
\def\bfr{{\bf r}}
\def\dtr{{\rm d}^3\bfr\,}
\def\bfx{{\bf x}}
\def\dtx{{\rm d}^3\bfx\,}
\def\dfx{{\rm d}^4 x\,}
\def\bfy{{\bf y}}
\def\dty{{\rm d}^3\bfy\,}
\def\dfy{{\rm d}^4 y\,}
\def\dfq{{{\rm d}^4 q\over (2\pi)^4}\,}
\def\dfk{{{\rm d}^4 k\over (2\pi)^4}\,}
\def\dfl{{{\rm d}^4 \ell\over (2\pi)^4}\,}
\def\dtq{{{\rm d}^3 {\bf q}\over (2\pi)^3}\,}
\def\dtk{{{\rm d}^3 {\bf k}\over (2\pi)^3}\,}
\def\dtl{{{\rm d}^3 {\bfl}\over (2\pi)^3}\,}
\def\dt{{\rm d}t\,}
\def\frac#1#2{{\textstyle{#1\over#2}}}
\def\darr#1{\raise1.5ex\hbox{$\leftrightarrow$}\mkern-16.5mu #1}
\def\){\right)}
\def\({\left( }
\def\]{\right] }
\def\[{\left[ }
\def\si{{}^1\kern-.14em S_0}
\def\siii{{}^3\kern-.14em S_1}
\def\diii{{}^3\kern-.14em D_1}
\def\dtwiii{{}^3\kern-.14em D_2}
\def\dthiii{{}^3\kern-.14em D_3}
\def\pziii{{}^3\kern-.14em P_0}
\def\poiii{{}^3\kern-.14em P_1}
\def\ptiii{{}^3\kern-.14em P_2}
\def\ipi{{}^1\kern-.14em P_1}
\def\idii{{}^1\kern-.14em D_2}
\def\fm{{\rm\ fm}}
\def\MeV{{\rm\ MeV}}
\def\CA{{\cal A}}
\def\Czzm{ {\cal A}_{-1[00]} }
\def\Cttm{{\cal A}_{-1[22]} }
\def\Ctzm{{\cal A}_{-1[20]} }
\def\Cztm{ {\cal A}_{-1[02]} }
\def\Czzz{{\cal A}_{0[00]} }
\def\Cttz{ {\cal A}_{0[22]} }
\def\Ctzz{{\cal A}_{0[20]} }
\def\Cztz{{\cal A}_{0[02]} }

\def\Ames{ A }  

\newcommand{\eqn}[1]{\label{eq:#1}}
\newcommand{\refeq}[1]{(\ref{eq:#1})}
\newcommand{\eq}{eq.~\refeq}
\newcommand{\eqs}[2]{eqs.~(\ref{eq:#1}-\ref{eq:#2})}
\newcommand{\eqsii}[2]{eqs.~(\ref{eq:#1}, \ref{eq:#2})}
\newcommand{\Eq}{Eq.~\refeq}
\newcommand{\Eqs}{Eqs.~\refeq}

\def\Journal#1#2#3#4{{#1} {\bf #2}, #3 (#4)}

\def\NCA{\em Nuovo Cimento}
\def\NIM{\em Nucl. Instrum. Methods}
\def\NIMA{{\em Nucl. Instrum. Methods} A}
\def\NPB{{\em Nucl. Phys.} B}
\def\NPA{{\em Nucl. Phys.} A}
\def\PLB{{\em Phys. Lett.} B}
\def\PRL{\em Phys. Rev. Lett.}
\def\PRD{{\em Phys. Rev.} D}
\def\PRC{{\em Phys. Rev.} C}
\def\PRA{{\em Phys. Rev.} A}
\def\ZPC{{\em Z. Phys.} C}
\def\SJP{{\em Sov. Phys. JETP}}

\def\FBS{{\em Few Body Systems Suppl.}}
\def\IJMP{{\em Int. J. Mod. Phys.} A}
\def\UJP{{\em Ukr. J. of Phys.}}


\def\spol{\alpha_{E0}}
\def\qpol{\alpha_{E2}}
\def\Mspol{\beta_{M0}}
\def\Mqpol{\beta_{M2}}


\preprint{\vbox{
\hbox{ NT@UW-98-16}
\hbox{ DUKE-TH-98-163}
}}
\bigskip
\bigskip

\title{The Polarizability of the Deuteron}
\author{Jiunn-Wei Chen,\footnote{\tt jwchen@phys.washington.edu},
  Harald W. Grie{\ss}hammer\footnote{\tt hgrie@phys.washington.edu},
  Martin J. Savage\footnote{\tt savage@phys.washington.edu} and Roxanne
  P. Springer\footnote{\rm  On leave from the Department of Physics, 
  Duke University, Durham NC 27708.
    \ \tt rps@redhook.phys.washington.edu.}}
\address{Department of Physics, University of Washington, Seattle, 
WA 98195-1560, USA }
\maketitle

\begin{abstract}
The scalar and tensor polarizabilities of the deuteron are calculated
using the recently developed effective field theory that describes 
nucleon-nucleon interactions.
Leading and next-to-leading order contributions in the perturbative 
expansion predict 
a scalar electric polarizability of
$\spol=0.595\ {\rm fm}^3$.
The tensor electric polarizability receives  contributions
starting at next-to-leading order from 
the exchange of a single potential pion
and is found to be 
$\qpol=-0.062\ {\rm fm}^3$.
We compute the leading contributions to the scalar and tensor
magnetic polarizabilities, finding
$\Mspol = 0.067 \ {\rm fm}^3$ and
$\Mqpol = 0.195 \ {\rm fm}^3$.

\end{abstract}
\vskip 1in

%
%
%
%
\vfill\eject


\section{Introduction}

Efforts to develop a systematic treatment of nucleon-nucleon interactions
\cite{Weinberg1,KoMany,Parka,KSWa,CoKoM,DBK,cohena,Fria,Sa96,LMa,GPLa,Adhik,RBMa,Bvk,aleph,Parkb,Gegelia,steelea,KSW} 
have culminated in an effective field theory with consistent 
power counting\cite{KSW}.
The leading contribution to two nucleons scattering in an
S-wave comes from local four-nucleon operators.  
Contributions from pion exchanges and from higher derivative operators
are suppressed by
additional powers of the external nucleon momentum and by powers of the light
quark masses.
The technique successfully describes the $NN$ scattering phase shifts up to
center-of-mass momenta of ${\bf p} \sim 300\ {\rm MeV}$ per nucleon\cite{KSW}
in all partial waves.

To accommodate the unnaturally large scattering lengths in S-wave
nucleon-nucleon scattering,
fine-tuning is required, which in turn 
can complicate power counting in the effective field 
theory.
Dimensional regularization with power divergence subtraction (PDS), described in
\cite{KSW}, provides a consistent power counting scheme.
Since the deuteron is the lightest nucleus and does not have 
irreducible forces
between three or more nucleons, it provides a unique laboratory for
studying the strong interactions.
Being bound by only $2.2\ {\rm MeV}$, the characteristic momentum
of the nucleons in the deuteron is $\sim 40\ {\rm MeV}$ and   should be
well described by the effective field theory which is valid below the scale
$\Lambda_{NN}\sim 300\ {\rm MeV}$\cite{KSW}.
The electromagnetic moments and  form factors of the deuteron have been
explored
with this new effective field theory\cite{KSW2}.
The charge radius and  form factor
receive
contributions from leading and next-to-leading (NLO)
orders in the expansion with the
theoretical result reproducing the measured values
within the uncertainty coming from the omission of higher order terms.
The magnetic moment and  form factor receive contributions at leading 
order from
only the nucleon magnetic moments.
At next-to-leading order  the deuteron magnetic moment
determines a combination of 
counterterms
that appear in the Lagrange density at this order.
The quadrupole moment first appears from the exchange of a single 
potential pion. Pion-pole contributions, multiple
potential pion exchange, and higher
dimension operators contribute only at higher orders in the expansion.

Unlike the electromagnetic form factors, the electric and magnetic
polarizabilities of an object are a direct measure of its ``deformation''
due to the presence of external electric and magnetic fields.
Extensive experimental and theoretical progress has been made in understanding
the
electric and magnetic  polarizabilities of the nucleon (for an overview see
\cite{Polwork}).
Chiral perturbation theory provides a systematic 
theoretical analysis in the single nucleon sector,
with one-loop pion
graphs dominating the electric polarizability,
e.g. \cite{BKMa,BS,Lvov,HolNa,Hola}.
The magnetic susceptibility, on the other hand, is dominated by the
$\Delta$-pole and has a significant uncertainty associated with it,
e.g. \cite{BSSa}.
Recently, the discussion has been extended to include ``generalized
polarizabilities,'' the amplitudes appropriate for interactions 
with electrons\cite{GLT,HHKS,DKMS}.

Theoretical understanding of the  polarizability of the deuteron has
been expressed in terms of 
meson exchange potential models\cite{FFa,PLH,LRa,LRb,MSZ,MKP,FPa,LLa,WWA,Khar}
(for an excellent discussion see \cite{FPa}).
In this work we present a model independent, analytic computation of the
electric polarizabilities of the deuteron to NLO
and the magnetic polarizabilities to leading order
in the effective
field theory describing nucleon-nucleon interactions.


\section{Effective Field Theory for Nucleon-Nucleon Interactions}

The terms in the effective Lagrange density describing the interactions
 between nucleons, pions,
and photons can be classified by the number of nucleon fields that 
appear.
It is convenient to write
\begin{equation}
{\cal L} = {\cal L}_0 + {\cal L}_1 + {\cal L}_2 + \ldots,
\end{equation}
where ${\cal L}_n$ contains $n$-body nucleon operators.

${\cal L}_0$ is constructed from the photon field $A^\mu = (A^0, {\bf A})$ and
the pion fields  which are incorporated into an $SU(2)$ matrix,
\begin{equation}
\Sigma = \exp\left({2i\Pi\over f}\right)\ \ \ ,\qquad \Pi = \left(\begin{array}{cc}
\pi^0/\sqrt{2} & \pi^+\\ \pi^- & -\pi^0/\sqrt{2}\end{array} \right)
\ \ \ \ ,
\end{equation}
where $f=132\ \MeV$ is the pion decay constant.  $\Sigma$ 
transforms under the global $SU(2)_L \times SU(2)_R$
chiral 
and $U(1)_{em}$ gauge symmetries as
\begin{equation}
\Sigma \rightarrow L\Sigma R^\dagger, 
\qquad \Sigma \rightarrow e^{i\alpha Q_{em}} \Sigma
e^{-i\alpha Q_{em}}
\ \ \ ,
\end{equation}
where $L\in SU(2)_L$, $R\in SU(2)_R$ and $Q_{em}$ is the charge matrix,
\begin{equation}
Q_{em} = \left(\begin{array}{cc}
1 & 0\\
0 & 0\end{array}\right)
\ \ \ .
\end{equation}
The part of the Lagrange density without nucleon fields is
\begin{eqnarray}
{\cal L}_0 &&= {1\over 2} ({\bf E}^2 - {\bf B}^2)
\ +\  {f^2\over 8} \Tr D_\mu \Sigma
D^\mu \Sigma^\dagger
\ +\  {f^2\over 4} \lambda \Tr m_q (\Sigma + \Sigma^\dagger) \ +\  \ldots
\ \ \ \ .
\end{eqnarray}
The ellipsis denote operators with more covariant derivatives $D_\mu$,
insertions of the quark mass matrix $m_q = {\rm diag} (m_u, m_d)$,
or factors of the
electric and magnetic fields.  
The parameter $\lambda$ has dimensions of mass and $m_\pi^2 = \lambda (m_u +
m_d)$.
Acting on $\Sigma$, the covariant derivative is
\begin{equation}
D_\mu \Sigma = \partial_\mu \Sigma + ie [Q_{em},\Sigma] A_\mu
\ \ \ .
\end{equation}

When describing pion-nucleon interactions, it is convenient to introduce the
field $\xi = \exp\left(i \Pi/f\right) = \sqrt{\Sigma}$.  
Under $SU(2)_L \times SU(2)_R$ this
transformations as
\begin{equation}
\xi \rightarrow L\xi U^\dagger = U\xi R^\dagger,
\end{equation}
where $U$ is a complicated nonlinear function of $L,R$, and the pion fields.
Since $U$ depends on the pion fields it has spacetime dependence.
The nucleon fields are introduced in a doublet of spin $1/2$ fields
\begin{equation}
N = \left({p\atop n}\right)
\end{equation}
that transforms under the chiral $SU(2)_L \times SU(2)_R$ symmetry as
$N \rightarrow UN$ and under the $U(1)_{em}$ gauge transformation as $N \rightarrow
e^{i\alpha Q_{em}} N$.
Acting on nucleon fields, the covariant derivative is
\begin{equation}
D_\mu N = (\partial_\mu + V_\mu + ie Q_{em}A_\mu )N \, \, ,
\end{equation}
where
\begin{eqnarray}
V_\mu &&= {1\over 2} (\xi D_\mu \xi^\dagger + \xi^\dagger D_\mu \xi)\nonumber \\
&& = {1\over 2} (\xi \partial_\mu \xi^\dagger + \xi^\dagger \partial_\mu\xi + ie
A_\mu (\xi^\dagger Q \xi - \xi Q \xi^\dagger)) \, \, .
\end{eqnarray}
The covariant derivative of $N$ transforms in the same way as $N$ under
$SU(2)_L \times SU(2)_R$ transformations (i.e. $D_\mu N \rightarrow U D_\mu
N$) and under $U(1)$ gauge transformations
(i.e. $D_\mu N \rightarrow e^{i\alpha Q_{em}} D_\mu N$).

The one-body terms in the Lagrange density are
\begin{eqnarray}
{\cal L}_1 & = & N^\dagger \left(i D_0 + {{\bf D}^2\over 2M_N}\right) N 
- {ig_A\over 2} N^\dagger {\bf \sigma} \cdot 
(\xi {\bf D} \xi^\dagger - \xi^{\dagger} {\bf D} \xi)
N\nonumber \\
& + &  {e\over 2M_N} N^\dagger
\left( \kappa_0 + {\kappa_1\over 2} [\xi^\dagger \tau^3\xi
  + \xi \tau^3 \xi^\dagger]\right) {\bf \sigma} \cdot {\bf B} N
\nonumber\\
& + &
2\pi \alpha_E^{(N0)} N^\dagger N {\bf E}^2\ +\ 2\pi \alpha_E^{(N1)} N^\dagger
\tau^3 N {\bf E}^2
\ +\ 
2\pi \beta_M^{(N0)} N^\dagger N {\bf B}^2\ +\ 
2\pi \beta_M^{(N1)} N^\dagger \tau^3 N {\bf B}^2
+ \ldots,
\label{lagone}
\end{eqnarray}
where
$\kappa_0 = {1\over 2} (\kappa_p + \kappa_n)$ and
$\kappa_1 = {1\over 2} (\kappa_p - \kappa_n)$
are isoscalar and isovector nucleon magnetic moments in nuclear magnetons, with
\begin{eqnarray}
\kappa_p & =&  2.79285\ ,\qquad\kappa_n = - 1.91304 \, \, .
\end{eqnarray}
The isoscalar and isovector electric polarizabilities of the nucleon are
$\alpha_E^{(N0)}$ and $\alpha_E^{(N1)}$ while the corresponding magnetic
quantities are
$\beta_M^{(N0)}$ and $\beta_M^{(N1)}$.
Experimentally, it is found that
\begin{eqnarray}
  \alpha_p & = & (12.1\pm 0.8\pm 0.5)\times 10^{-4}\ {\rm fm}^3
  \ \ ,\ \ \beta_p = (2.1\mp 0.8\mp 0.5)\times 10^{-4}\ {\rm fm}^3
\ \ \ ,
\end{eqnarray}
for the proton \cite{ppola,ppolb}
while the two  measurements of the neutron electric  polarizability 
$\alpha_n = 12.0\pm1.5\pm 2.0$ \cite{npola}\ and
$\alpha_n = 0\pm 5$ \cite{npolb} indicate a sizable uncertainty.

The two-body Lagrange density is 
\begin{eqnarray}
\CL_2 &=& -\left(C_0^{(\siii)}+ D_2^{(\siii)} \lambda\Tr m_q\right))
(N^T P_i N)^\dagger(N^T P_i N)
\nonumber\\
 & + & {C_2^{(\siii)}\over 8}
\left[(N^T P_i N)^\dagger
\left(N^T \left[ P_i \overrightarrow {\bf D}^2 +\overleftarrow {\bf D}^2 P_i
    - 2 \overleftarrow {\bf D} P_i \overrightarrow {\bf D} \right] N\right)
 +  h.c.\right]
\nonumber\\
&& + e L_1 (N^\dagger {\bf \sigma} \cdot {\bf B}N) (N^\dagger N) +
e L_2 (N^\dagger {\bf \sigma} \cdot {\bf B} \tau^a N)
(N^\dagger \tau^a N)
\nonumber\\
& + & 2\pi \alpha_4  (N^T P_i N)^\dagger(N^T P_i N) {\bf E}^2
\ +\ 2\pi \beta_4  (N^T P_i N)^\dagger(N^T P_i N) {\bf B}^2
+ \ldots,
\label{lagtwo}
\end{eqnarray}
where $P_i$ is the spin-isospin projector for the spin-triplet channel
appropriate for the deuteron
\begin{eqnarray}
P_i \equiv {1\over \sqrt{8}} \sigma_2\sigma_i\tau_2
\ \ \ , 
\qquad \Tr P_i^\dagger P_j ={1\over 2} \delta_{ij}
\ \ \ .
\end{eqnarray}
The $\sigma$ matrices act on the nucleon spin indices,
while the $\tau$ matrices act on isospin indices.
The local operators responsible for $S-D$
mixing do not contribute at either leading or NLO.
The $C_{0}^{(\siii)}$, $C_2^{(\siii)}$ and $D_2^{(\siii)}$ coefficients
are determined from 
$NN$ scattering to be
\begin{eqnarray}
C_0^{(\siii)}(m_\pi) =-5.51\fm^2\ ,\ 
D_2^{(\siii)}(m_\pi) =1.32\fm^4\ ,\ 
C_2^{(\siii)}(m_\pi) =9.91\fm^4
\ ,
\label{eq:numfitc}
\end{eqnarray}
where we have chosen to renormalize the theory at a scale $\mu=m_\pi$
in the PDS scheme\cite{KSW}.
A  linear combination of the  coefficients $L_{1,2}$ 
contribute to the magnetic
moment of the deuteron, but neither they nor the
coefficients $\alpha_4$ and $\beta_4$ contribute to the deuteron
polarizability at the order to which we are working.
The four individual nucleon polarizability counterterms 
terms also do not contribute
at the order to which we are working.
In eq.~(\ref{lagtwo}) we have only shown the leading terms
of the expansion in meson fields, namely the terms we need for our leading
plus NLO calculation.

Since we are working with a field theory, no
ambiguities arise from how we choose to define the interpolating fields.
We can consistently neglect  all operators that vanish by the
equations of motion \cite{Politzer,Arzt}.
An operator that vanishes by the equations of motion makes a contribution
to an observable that has exactly the same form as higher dimension
operators that are present in the theory.
Further, such operators can be removed
by field redefinitions, and therefore it is consistent to work with a Lagrange
density
that does not contain operators that vanish by the equations of motion.

Another point of interest is that we do not need to include the $\Delta$ as an
explicit degree of freedom in the theory.
In order to have a theory that is well
defined for processes involving momenta up to $\sim 1\ {\rm GeV}$, the
$\Delta$ must be included as a dynamical object\cite{JMa}.
However, the power counting for the effective field theory 
describing the nucleon-nucleon
interaction outlined in this section is limited to momenta less than
${\bf  p}\sim \Lambda_{NN}\sim 300\ {\rm MeV}$.
The momentum scale making the 
dominant contribution to graphs involving the $\Delta$ is approximately
${\bf p}\sim\sqrt{M_N (M_\Delta-M_N)}\sim 500\ {\rm MeV}$, higher than
$\Lambda_{NN}$.
Therefore, the $\Delta$ is not included as a dynamical object, but its effects
are included in the coefficients of the local operators
(as are the effects of all particles not included as dynamical degrees of freedom).


\section{Computing the Polarizabilities of the Deuteron}

The Lagrange density described in the previous section in terms of nucleon
field operators matches onto an effective theory describing the
dynamics of the deuteron field
\begin{eqnarray}
  {\cal L}_D & = &
  {\cal D}^\dagger_a \left(i D_0 + {{\bf D}^2\over 2 M_D}\right) {\cal D}^a
  \ -\
  i\mu_D\ \epsilon^{abc}  {\cal D}^\dagger_a {\cal D}_b {\bf B}_c
  \nonumber\\
 & + & 
  2\pi\alpha_{E0}\ {\cal D}^\dagger_a {\cal D}^a {\bf E}^2
  \ +\
  2\pi\beta_{M0}\ {\cal D}^\dagger_a {\cal D}^a {\bf B}^2
  \nonumber\\
  & + &
  2\pi\alpha_{E2}\ \left[{\cal D}^\dagger_a {\cal D}_b +{\cal D}^\dagger_b
    {\cal D}_a - {2\over 3}\delta_{ab}  {\cal D}^\dagger_c {\cal D}^c\right]
    {\bf E}^a {\bf E}^b
  \nonumber\\
  & + &
  2\pi\beta_{M2}\ \left[{\cal D}^\dagger_a {\cal D}_b +{\cal D}^\dagger_b
    {\cal D}_a - {2\over 3}\delta_{ab}  {\cal D}^\dagger_c {\cal D}^c\right]
   {\bf B}^a {\bf B}^b  
\ \ \ .
\label{lagDeut}
\end{eqnarray}
${\cal D}^a$ is an operator that annihilates a deuteron, and its spin index
takes values $a=1,2,3$.
The covariant derivative acting on the deuteron field is
$D_\mu = \partial_\mu + i e Q A_\mu$.
The coefficient $\mu_D$ is the deuteron magnetic moment and has been determined
from eqs.(\ref{lagone}) and (\ref{lagtwo}) to NLO in \cite{KSW2}.
The scalar electric polarizability of the deuteron is $\alpha_{E0}$, and the
scalar
magnetic polarizability is $\beta_{M0}$.   
These operators give rise to
interactions that are independent of the alignment  of the
deuteron with respect to an applied electromagnetic field.
The tensor electric polarizability of the deuteron is $\alpha_{E2}$,
and the tensor
magnetic polarizability is $\beta_{M2}$.
These operators give rise to interactions that depend upon the alignment of
the deuteron with respect to an applied electromagnetic field.
In order to compute $\alpha_{E0,E2}$ and $\beta_{M0,M2}$ we will use the
formalism developed in \cite{KSW2}.   
The polarizabilities $\alpha$ and $\beta$ each have perturbative expansions in
powers of 
$Q=m_\pi/\Lambda_{NN}\ \sim\ \sqrt{M_N B}/\Lambda_{NN}$, 
where $B$ is the deuteron
binding energy.
We  denote the contribution from a given order by a superscript,
e.g. the scalar electric polarizability
\begin{eqnarray}
  \alpha_{E0} & = & \alpha_{E0}^{(-4)}\ +\ \alpha_{E0}^{(-3)}\ +\ ..
\ \ \ ,
\end{eqnarray}
and similarly for the other polarizabilities.

In matching onto the deuteron effective Lagrange density in eq.~(\ref{lagDeut})
from the Lagrange density describing nucleon dynamics,
eqs.~(\ref{lagone}) and (\ref{lagtwo}),
we will recover the coefficient of each operator order by order in the 
$Q$ expansion, including the 
coefficient of  the operator $ {\cal D}^\dagger_a {\bf D}^2 {\cal D}^a$.
At leading order in the loop expansion, we find this operator to have
a coefficient
${1\over 4 M_N}$, while at next-to-next-to-leading order (NNLO)
it
will have a coefficient
${1\over 4 M_N} + {B\over 8 M_N^2}$.
When determined to all orders it will sum
to
${1\over 2( 2 M_N-B)} = {1\over 2 M_D}$, which we have written in
eq.~(\ref{lagDeut}). 
However, it is important to realize that we will only recover interactions
coming from this operator order by order in the expansion.

\begin{figure}[t]
\centerline{{\epsfxsize=1.8in \epsfbox{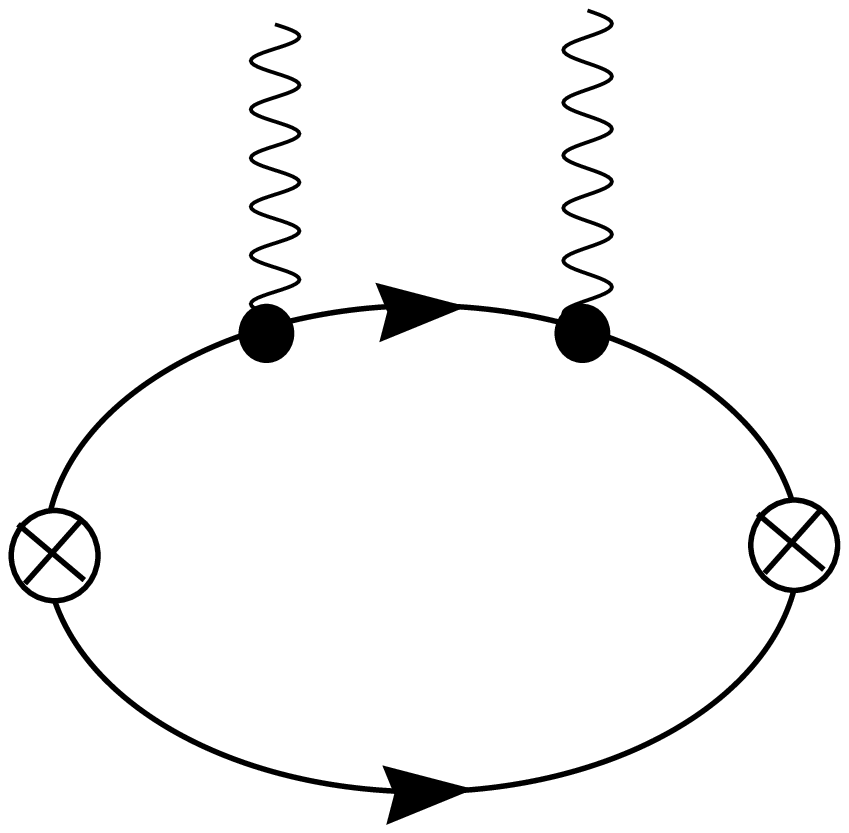}}\hskip 0.4in
            {\epsfxsize=4.0in \epsfbox{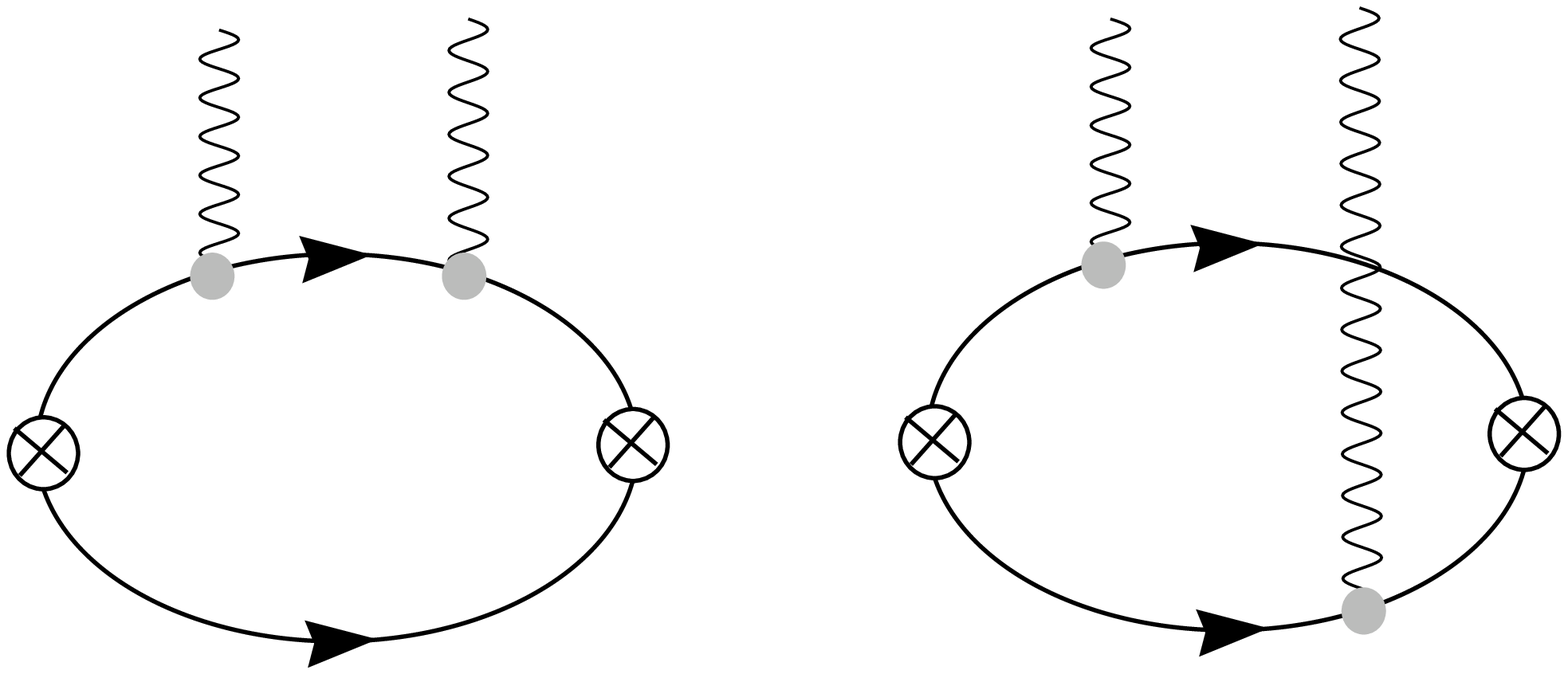}} }
\vskip 0.1in
\centerline{ {\epsfxsize=6.5in \epsfbox{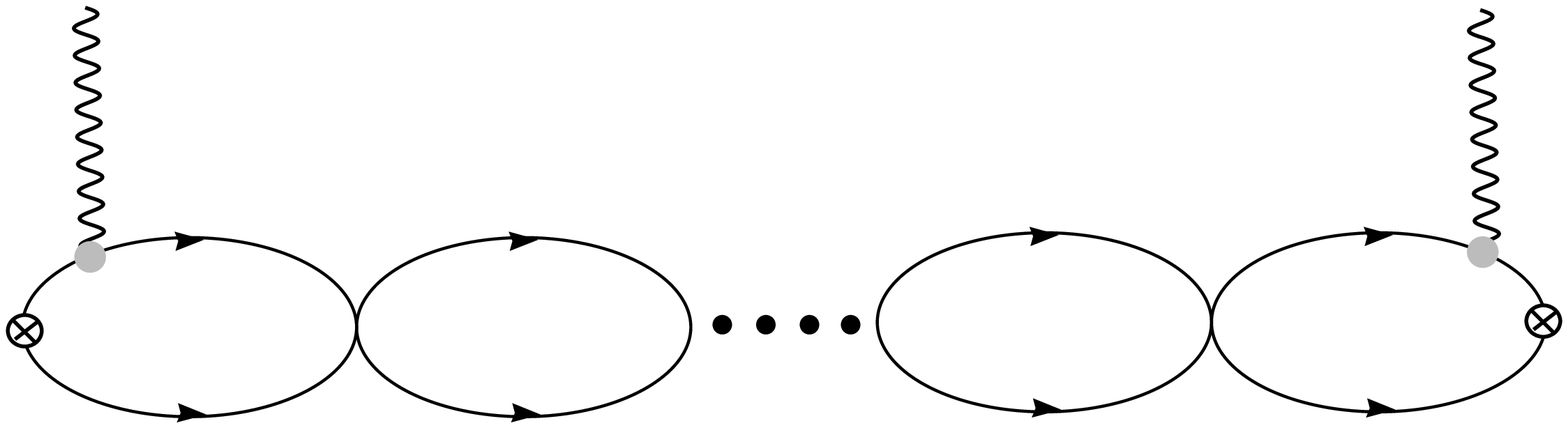}}}
\noindent
\caption{\it Leading order contributions to the deuteron
  polarizabilities.
  The crossed circles denote operators that create or annihilate
  two nucleons with
  the quantum numbers of the deuteron.
  The dark solid circles correspond to the photon coupling via 
  the nucleon kinetic energy operator (minimal coupling),
while the light solid circles denote the nucleon magnetic moment operator.
The solid lines are nucleons.
Only the graph with the nucleons minimally 
coupled to the electromagnetic field 
contributes to the electric polaribility.
  The photon crossed graphs are not shown.
  The bubble chain arises from insertions of the four nucleon operator
  with coefficient $C_0^{(\si)}$.
}
\label{fig:Pollead}
\vskip .2in
\end{figure}

\begin{figure}[t]
\centerline{\epsfxsize=2.0in \epsfbox{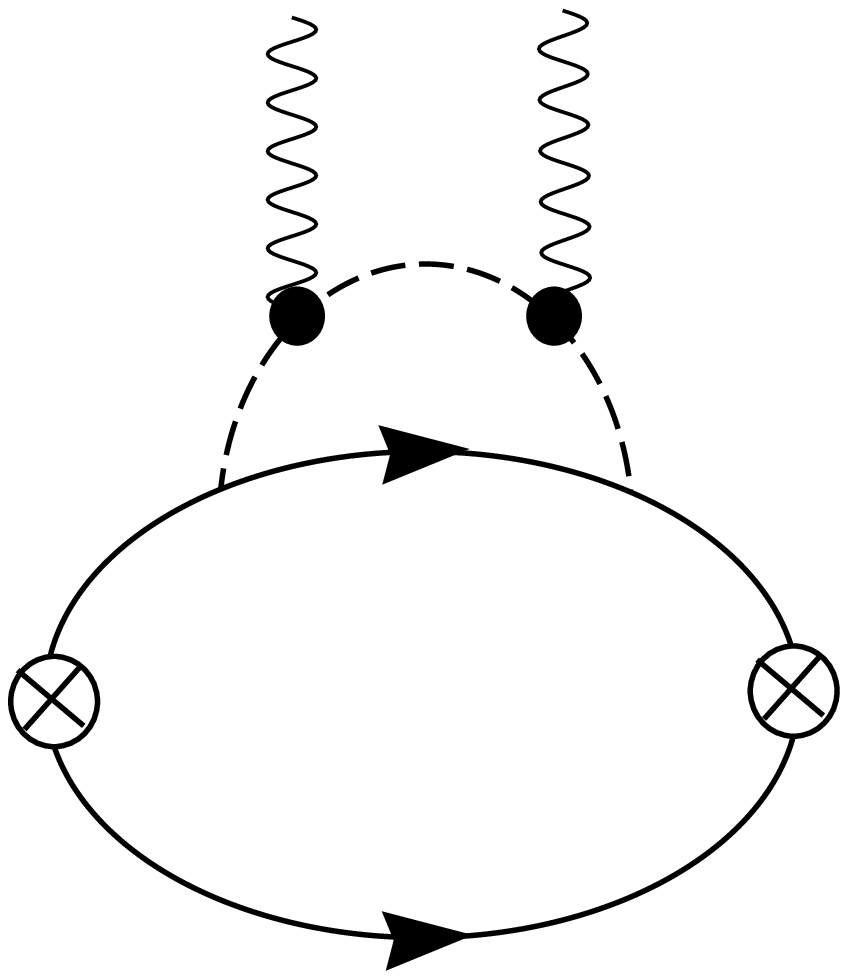}}
\noindent
\caption{\it A contribution to the deuteron electric
  polarizability from 
  a graph that also contributes to 
  the
  polarizability of the nucleon. 
  The crossed circles denote operators that create or annihilate
  two nucleons with
  the quantum numbers of the deuteron.
  The solid circles correspond to the photon coupling via 
  the nucleon kinetic energy operator (minimal coupling).
  Dashed lines
  are pions and solid lines are nucleons.
}
\label{fig:NucPol}
\vskip .2in
\end{figure}

 It is instructive to begin by power counting the contributions from operators
 that appear in the Lagrange density eqs.\ (\ref{lagone}) and (\ref{lagtwo}).
 Consider the
 single loop graph of Fig.~\ref{fig:Pollead}
 between a source that creates a spin
 triplet $NN$ pair and one that annihilates it, minimally coupling to two
 photons.  
In terms of the expansion
 parameter $Q$, a non-relativistic loop integral scales as $Q^5$,
 a non-relativistic nucleon propagator as $Q^{-2}$, and a gradient operator as
 $Q^1$.
In order to match onto the ${\bf E}^2$ or
${\bf B}^2$ operator in the deuteron Lagrange density (\ref{lagDeut}),
we must find the coefficient of $\omega^2$ or ${\bf k}^2$
by expanding 
the graph in powers of $\omega$ and ${\bf k}$,
respectively, where $(\omega,{\bf k})$ is the photon four-momentum. 
Therefore, this diagram will scale like
$Q^{-1}\sum_n\ \left({ M_N \omega , |{\bf k}|^2\over Q^2}\right)^n$.
Wave function renormalization introduces a factor of $Q$,
and therefore we find that the leading
contribution to the electric polarizabilities arising from this graph is of
order $Q^{-4}$. This power counting also shows that the $C_2(\mu)$ operator
and the exchange of a single potential pion contribute
at order $Q^{-3}$ after wave
function renormalization.
The magnetic polarizabilities receive contributions from all the graphs shown
in Fig.~\ref{fig:Pollead} at order $Q^{-2}$.
The four-nucleon polarizability
counterterms, $\alpha_4$ and $\beta_4$,
 appearing in (\ref{lagtwo}) contribute at
order $Q^{1}$, and can be safely neglected.  
Counterterms for the
individual nucleon polarizabilities appearing in (\ref{lagone}) contribute
at order $Q^0$.  
Meson loop corrections that are the dominant contribution
to the electric polarizability of the nucleon, for instance
Fig.~\ref{fig:NucPol},
appear at order $Q^{-1}$,
 three orders higher in the expansion than the leading order
contributions.
Therefore, it is probable that the polarizabilities of the 
individual nucleons will not be
extracted from the polarizabilities of the deuteron.

We will compute the leading and NLO contributions to the scalar and
tensor electric polarizability of the deuteron,
and the leading contributions to the magnetic scalar and tensor
polarizabilities.
The loop graphs in Fig.~\ref{fig:Pollead}
give
the leading contribution to the deuteron polarizabilities,
which after wave function renormalization are
\begin{eqnarray}
  \alpha_{E0}^{(-4)} & = & {\alpha M_N\over 32 \gamma^4}
  \nonumber\\
\alpha_{E2}^{(-4)} & = & 0
  \nonumber\\
\beta_{M0}^{(-2)}  & = & {\alpha\over 2 M_N}
  \left[ -{1\over 16\gamma^2} \ +\
    {2(\kappa^{(0)})^2 +(\kappa^{(1)})^2 \over 3\gamma^2}
    \ +\
    { (\kappa^{(1)})^2\over 6\pi}{M_N\over \gamma} {\cal A}_{-1}^{(\si)}(-B)
    \right]
  \nonumber\\
\beta_{M2}^{(-2)} & = & -{\alpha\over 2M_N}
\left[ 
  {(\kappa^{(0)})^2 -(\kappa^{(1)})^2 \over \gamma^2}
  \ -\
    { (\kappa^{(1)})^2 M_N \over 2\pi\gamma} {\cal A}_{-1}^{(\si)}(-B) 
    \right] 
\ \ \ ,
\label{eq:allead}
\end{eqnarray}
where $\gamma=\sqrt{M_N B}$ and $\alpha=1/137$ the fine structure constant.
The scattering amplitude in the $\si$ channel 
for a centre of mass energy $E$ is 
\begin{eqnarray}
{\cal A}_{-1}^{(\si)}(E) & = & {- C_0^{(\si)} 
\over 
1 +  C_0^{(\si)} {M_N\over 4\pi}(\mu-\sqrt{-M_N E-i\varepsilon})}
\ \ \ ,
\end{eqnarray}
which is of order $Q^{-1}$.
The coefficient $C_0^{(\si)}$ has been determined from 
nucleon-nucleon scattering in the $\si$ channel to be \cite{KSW}
$C_0^{(\si)} = -3.34\ {\rm fm^2}$.


\begin{figure}[t]
\centerline{\epsfxsize=4.0in \epsfbox{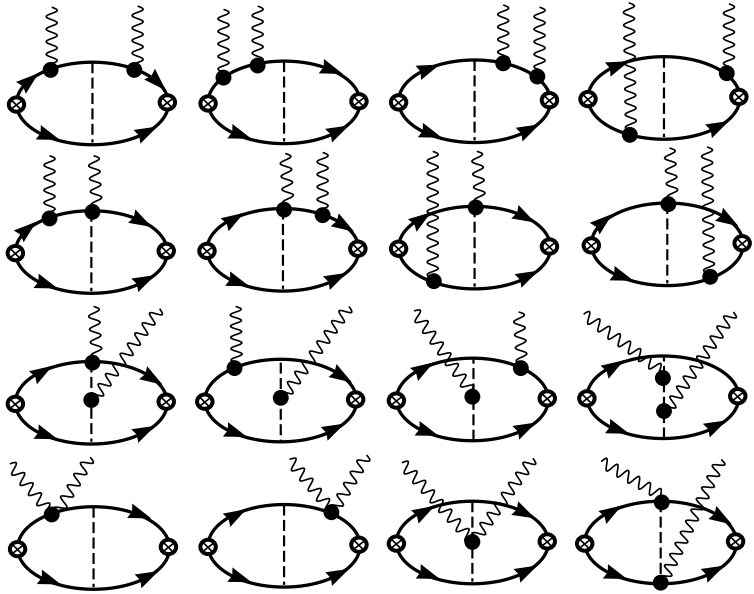}}
\noindent
\caption{\it Graphs from potential pion exchange that contribute to 
the deuteron polarizabilities at NLO.  
  The crossed circles denote operators that create or annihilate
  two nucleons with
  the quantum numbers of the deuteron.
  The solid circles correspond to the photon coupling via 
  the nucleon or meson kinetic energy operator (minimal coupling)
  or from the gauged axial coupling to the meson field.  Dashed lines
  are mesons and solid lines are nucleons.
  Photon crossed graphs are not shown.
  }
\label{fig:Polsubpi}
\vskip .2in
\end{figure}
%
%

\begin{figure}[t]
\centerline{\epsfxsize=4.0in \epsfbox{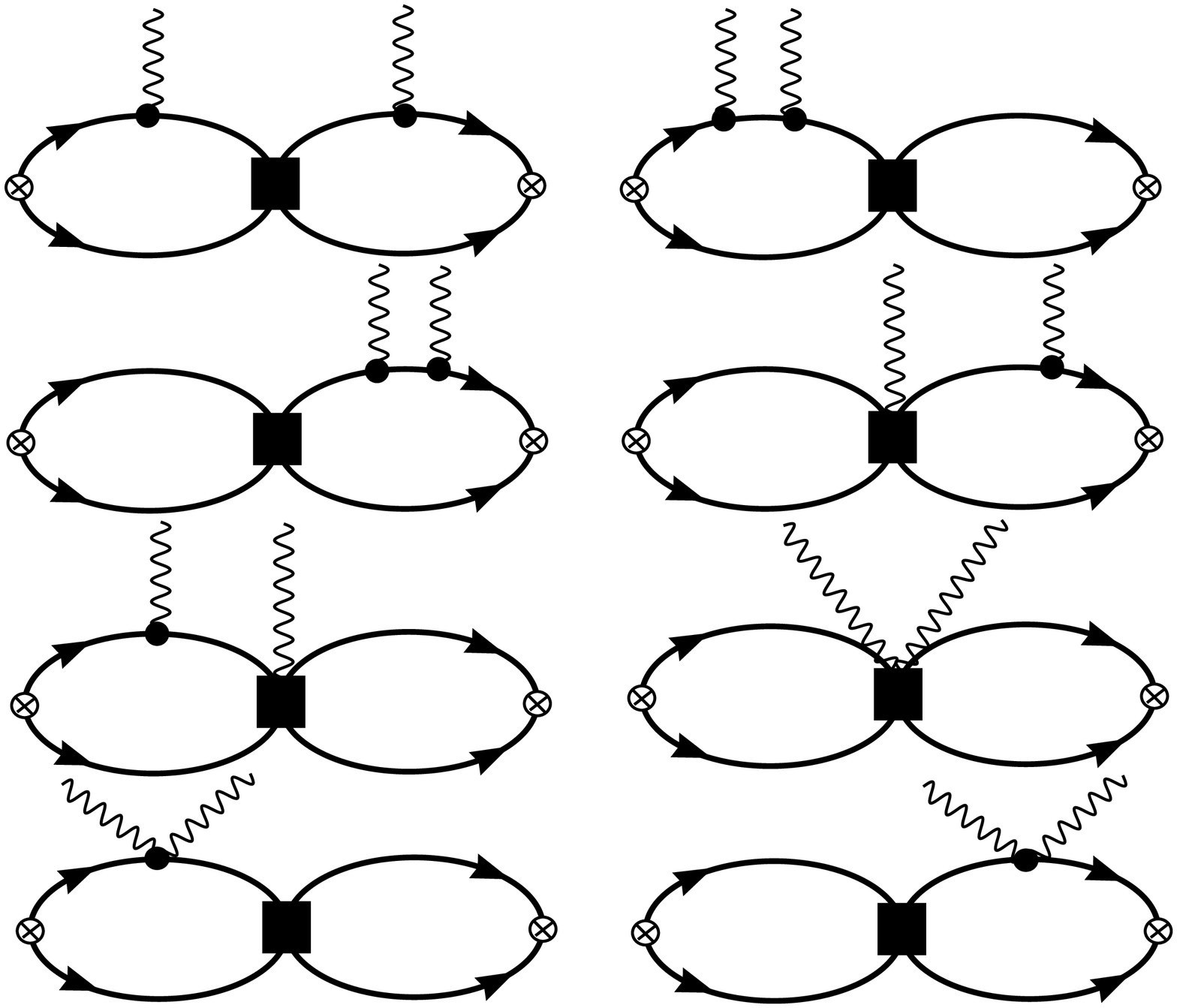}}
\noindent
\caption{\it Graphs from insertions of the operator 
with coefficient $C_2(\mu)$ that contribute to 
the deuteron polarizabilities at NLO.  
  The crossed circles denote operators that create or annihilate 
  two nucleons with
  the quantum numbers of the deuteron.
  The solid circles correspond to the photon coupling via 
  the nucleon kinetic energy operator (minimal coupling) while
  the solid square denotes the $C_2(\mu)$ operator.
  The solid lines are nucleons.
  Photon cross graphs are not shown.
  }
\label{fig:PolsubC}
\vskip .2in
\end{figure}

At NLO there are contributions from the
exchange of a single potential pion, 
Fig.~\ref{fig:Polsubpi}, and from the operator with 
coefficient $C_2(\mu)$,
Fig.~\ref{fig:PolsubC}.
The operator with coefficient $D_2(\mu)$ does not contribute to the
polarizability of the deuteron.
We find that at order $Q^{-3}$ 
\begin{eqnarray}
  \alpha_{E0}^{(-3)} & = & {\alpha M_N^2\over 64\pi\gamma^3} 
  C_2(\mu) (\mu-\gamma)^2
\ +\
{\alpha g_A^2 M_N^2\over 384\pi f^2}\ 
  { m_\pi^2 ( 3 m_\pi^2 + 16 m_\pi\gamma + 24\gamma^2)\over
      \gamma^3 (m_\pi+2 \gamma)^4}
    \nonumber\\
\alpha_{E2}^{(-3)} & = & - {\alpha g_A^2 M_N^2\over 80\pi f^2} \ 
  { 2m_\pi^3+11 m_\pi^2\gamma + 16 m_\pi\gamma^2 + 8\gamma^3\over \gamma^2
    (m_\pi+2\gamma)^4}
  \ \ \ \ .
  \label{eq:alsub}
\end{eqnarray}
The leading contribution to the tensor polarizability of the
deuteron $\alpha_{E2}$
comes
from the exchange of a potential pion.
The explicit renormalization scale dependence that appears in
eq.~(\ref{eq:alsub}) is compensated by the $\mu$ dependence of the
coefficient $C_2(\mu)$, which scales $\sim 1/\mu^2$\cite{KSW}.

Numerically, we find that the electric polarizabilities are 
\begin{eqnarray}
  \alpha_{E0} & = &  \alpha_{E0}^{(-4)}\ +\ \alpha_{E0}^{(-3)}\ +\ ...
  \nonumber\\
  & = &  0.386\ {\rm fm}^3\ +\  ( 0.153 +  0.057)\ {\rm fm}^3\ +\ ...
  \nonumber\\
  & = &  0.595\ {\rm fm}^3\ +\ ...
\ \ \ \ ,
\label{eq:snum}
\end{eqnarray}
where the dots denote contributions higher order in the expansion.
The first term in parenthesis on the second line of eq.~(\ref{eq:snum})
comes from
the  $C_2(\mu)$ operator while the second term arises from the exchange of a
single potential pion.
The convergence of the expansion for $  \alpha_{E0}$ appears to be
approximately the same as it is for the static electromagnetic moments
\cite{KSW2}, with each order being suppressed by between $1/3$ and $1/2$,
consistent with the $\Lambda_{NN}$ expansion\cite{KSW}.
We expect that the uncertainty in this numerical value is roughly
$\Delta\alpha_{E0}\approx 0.1\ {\rm fm}^3$
from an estimate of the next order contribution.
For the tensor polarizability we find
\begin{eqnarray}
  \alpha_{E2} & = &  \alpha_{E2}^{(-4)}\ +\ \alpha_{E2}^{(-3)}\ +\ ...
  \nonumber\\
& = & -0.062\ {\rm fm}^3\ +\ ...
\ \ \ ,
\label{eq:qnum}
\end{eqnarray}
where we recall $\alpha_{E2}^{(-4)}=0$.
The fractional uncertainty in  $\alpha_{E2}$ is much greater than  that 
for $\alpha_{E0}$
as it has a vanishing leading order contribution, and we naively estimate an
uncertainty of $\Delta\alpha_{E2}\sim 0.03\ {\rm fm}^3$.
The leading contribution to the magnetic polarizabilities are
\begin{eqnarray}
  \beta_{M0}^{(-2)} & = & 0.067\ {\rm fm}^3\ \ \ \ ,
  \nonumber\\
  \beta_{M2}^{(-2)} & = & 0.195\ {\rm fm}^3
  \ \ \ \ .
\label{eq:magp}
\end{eqnarray}
The  large values for the magnetic polarizabilities
come from  the isovector magnetic moment of the nucleon,
$\kappa_1$.

It is informative to decompose these interactions into the polarizabilities
of the individual magnetic substates of the deuteron.
The electric polarizability
of the $m=\pm 1$ states of the deuteron is
$\alpha_{E0} - {2\over  3}\alpha_{E2}$
while the polarizability of the $m=0$ state is
$\alpha_{E0} + {4\over  3}\alpha_{E2}$,
and similarly for the magnetic polarizabilities, we  find
\begin{eqnarray}
  \alpha_{E}^{\bf |m|=1} & = & 0.637\ {\rm fm}^3\ +\ ...
  \ \ \ ,\ \ \
  \beta_{M}^{\bf |m|=1}  =  -0.063\ {\rm fm}^3\ +\ ...
  \nonumber\\
  \alpha_{E}^{\bf m=0} & = &  0.511\ {\rm fm}^3\ +\ ...
  \ \ \ ,\ \ \
  \beta_{M}^{\bf m=0}  =  0.327\ {\rm fm}^3\ +\ ...
\ \ \ ,
\label{eq:mag}
\end{eqnarray}
where the ellipsis denotes higher order contributions.
Numerical evaluation of our analytic results 
agree within uncertainties with values for the
electric polarizabilities obtained
from potential models.
To a very high degree of precision potential models predict a scalar
electric polarizability of
$\alpha_{E0} = 0.6328\pm 0.0017\ {\rm fm}^3$\cite{FPa},
which is remarkably consistent with the ``zero-range'' limiting value of
$\alpha_{E0} = 0.632\pm 0.003\ {\rm fm}^3$\cite{FFa}.
Further, the calculations of \cite{Khar} find
$\alpha_{E}^{\bf |m|=1}  =  0.669\ {\rm fm}^3$
and 
$\alpha_{E}^{\bf m=0} =  0.555\ {\rm fm}^3$.
We expect to deviate from these values at higher orders in the
effective field theory expansion
as generally potential models do not properly describe chiral dynamics, pion
propagation, and relativistic effects.
The power of the effective field theory  formalism is that such effects can
be included systematically and the natural size of higher order terms is known.
Effective field theories will never match the precision of 
potential or other models because their reliance on {\it only}
the symmetries of the underlying theory and experimental results
necessarily makes them less restrictive.  However, the  fact that 
effective field theory descriptions provide a model independent treatment
makes them an important tool in the study of low energy processes.


\section{Conclusions}

We have performed a model independent
computation of the scalar and tensor polarizabilities of the deuteron
in the effective theory describing nucleon-nucleon interactions.
The scalar electric polarizability receives leading order contributions
while the tensor electric polarizability begins at NLO in the
power counting, and so is sizably smaller than the scalar polarizability.
The tensor electric polarizability arises only from the exchange of 
potential pions at NLO.
The leading contribution to the magnetic polarizability of the deuteron
comes from the magnetic moments of the individual nucleons in addition
to their minimal coupling to the electromagnetic field.
From this analysis we conclude that it is unlikely 
that the neutron polarizabilities will be extracted
from the deuteron polarizabilities.

We are encouraged that this effective field theory calculation
reproduces the results of potential model calculations.  
The benefits
of using an effective theory  are that   closed form
expressions have been obtained 
for the polarizabilities, and through the universal couplings
in the Lagrange density we can see how QCD relates a variety
of two nucleon processes in a systematic way.
Further, higher order terms  not considered in this work are parametrically 
smaller that those we have presented.
In subsequent work we will present the photon-deuteron scattering cross section
over a range of photon energies.

\vskip 0.5in

This work is supported in part by the U.S. Dept. of Energy under
Grants No. DE-FG03-97ER4014 and DE-FG02-96ER40945.

\end{document}